\documentclass[a4paper]{jpconf}
\usepackage{graphicx}
\begin{document}
\title{Microlensing towards the Magellanic Clouds and M31: is the quest for MACHOs still open?}

\author{Sebastiano Calchi Novati}

\address{Dipartimento di Fisica ``E. R. Caianiello'', Universit\`a di Salerno, 
Via Ponte don Melillo, 84084 Fisciano (SA), Italy \&
Istituto Internazionale per gli Alti Studi Scientifici (IIASS), Vietri Sul Mare (SA), Italy}

\ead{novati@sa.infn.it}

\begin{abstract}
Microlensing is the tool of choice for the search
and the analysis of compact halo objects (``MACHOs''), a still
viable class of dark matter candidates at the galactic scale.
Different analyses point towards an agreement
in excluding dark matter MACHOs of less than about $10^{-1}~\mathrm{M}_\odot$;
it remains however an ongoing debate for values in the mass range $(0.1-1)~\mathrm{M}_\odot$.
The more robust constraints, though not all in agreement, 
come from the observational campaigns towards the Magellanic Clouds
(the LMC and the SMC).
The analyses towards the nearby
galaxy of M31, in the so called ``pixel lensing'' regime,
have expanded the perspectives in this field of research.
In this contribution first we draw 
a critical view on recent results 
and then we focus on the pixel lensing analysis 
towards M31 of the PLAN collaboration.
\end{abstract}

\section{Introduction}

The original suggestion of Paczy\'{n}ski (1986)
of using microlensing as a tool to probe
compact halo objects as a dark matter candidate
at the galactic scale has given rise, in the meantime,
to a much larger range of applications of microlensing which
has become an efficient tool of research
for many different relevant astrophysical issues.
(The observational driving reason
being the relatively large rate of events
expected, and by now routinely observed, towards the Galactic bulge.)
A  partial list includes the study of the inner Galactic structure
(Moniez, 2010 for a review),
many aspects of stellar astrophysics (Gould, 2001) with in particular
analyses on the mass function towards the Bulge (Gould, 2000; Calchi~Novati et~al, 2008)
and the recent exciting result on an unbound or distant planetary mass population 
(Sumi et~al, 2011). Currently, however, the main field of application of microlensing
has become the search of exoplanets 
thanks to its ability to explore different part of the planetary
parameter space with respect to other techniques (Gaudi, 2010; Dominik, 2010 for recent reviews). 
Within this framework first results on the frequency of exoplanets 
have also been reported (Sumi et~al, 2010; Gould et~al, 2010).
Microlensing is indeed well suited to infer the statistical properties
of the underlying exoplanet population, still, an adequate
strategy to this purpose must be adopted (Dominik et al, 2010).

The search of MACHOs remains however an active field of research,
with observational campaigns being carried out to this purpose both towards
the Magellanic Clouds and M31. As more thoroughly discussed in the following sections,
the debate is currently centered on an hypothetical 
MACHO population in the mass range $(0.1-1)~\mathrm{M}_\odot$.
This is however the very same range of mass for
normal stars acting as lenses (we broadly refer
to this case as ``self lensing'', as opposed to MACHO lensing),
the lens mass being the driving physical parameter 
of the observed characteristics of microlensing light curves.
This coincidence may therefore be taken as an hint
of some bias in the analyses.
If not the case, however, the reported sizeable
fraction of MACHOs in this mass range (up to about $f=20\%$ in mass of the halo)
would represent a real astrophysical challenge certainly
worth more detailed analyses.

\section{The results towards the Magellanic Clouds}
The main results along this line of sight 
have been reported by the MACHO, the EROS
and the OGLE collaborations. We refer
to Moniez (2010) for a detailed discussion on the subject and a full list of references.
The relevant and definite result of all these observational
campaigns is that compact halo object \emph{are not} a major component
of the Galactic halo. However, some disagreement remains
to be clarified in some of the considered mass ranges. 
There is a full consensus to exclude, for a fraction in  mass
well below 10\%, MACHOs as viable dark matter candidates from about $10^{-7}~\mathrm{M}_\odot$
up to a mass below about $0.1~\mathrm{M}_\odot$.
This agreement then breaks down in the mass range $(0.1-1)~\mathrm{M}_\odot$.
For larger values of the MACHO mass the smaller number of expected candidates makes the
constraints less tight (although, in a re-analysis of the OGLE-II
and OGLE-III data, Calchi~Novati and Mancini (2011) obtained
the stronger bounds to date by microlensing in this mass range).
More in particular, the MACHO collaboration (Alcock et~al, 2000a) 
reported an evidence of a signal, $f\sim 20\%$ 
(in the hypothesis of a ``standard'' pseudo-isothermal spherical
density distribution for the halo\footnote{As it has been already
observed, eg Bennett (2005), the fact that MACHOs, if any,
are not the main component of dark matter haloes,
also implies that they may not follow its spatial
distribution. In turn, this imply that 
the actual mass fraction related to the
observed events, if really to be attributed to MACHO lensing,
may in fact be smaller than that evaluated 
for a standard halo model.}) for 
about $0.5~\mathrm{M}_\odot$ MACHOs out of observations towards the LMC. 
Two critical aspects of this analysis are worth being recalled:
MACHO monitored mainly  the central LMC region,
where one can expect a larger contamination by the LMC self-lensing
signal; MACHO considered as viable sources also rather faint stars
and this complicates the analysis of the detection efficiency
and therefore the related aspect of the estimate of the expected signal.
The results of the MACHO collaboration, substantially confirmed in 
a more recent analysis by Bennett (2005), 
have then been  challenged by the EROS collaboration (Tisserand et~al, 2007) and more
recently by the analyses of the OGLE collaboration whose
OGLE-II and OGLE-III campaigns towards both
the LMC and the SMC have been discussed in a recent series of papers
by Wyrzykowski et~al (2009, 2010, 2011a,b).
Taken at face value, all these last analyses indicate that self lensing alone is able
to explain the observed rate of events, clearly at odds with
the result of the MACHO collaboration.

Comparing to the case of M31, to be discussed
in the next Section, it is worth recalling
a few points specific to the observations
towards the Magellanic Clouds
(in addition to the different microlensing regime
one enters when looking at the
more distant sources in M31).
First, an intrinsic bias in the results
may be introduced by the implicit
underlying, though not probed, assumption
that the single line of sight probed in this case
through the Galactic halo is 
representative of this full component.
This may not be the case for instance
because of tidal interactions 
with the Galaxy and/or local halo clumpiness. 
Second, the expected rate by ``self lensing''
is small (with a sizeable fraction of lenses
to be expected from the Milky Way disc,
as recently pointed out by Calchi~Novati and Mancini, 2011),
both in absolute number, at least for the 
reported observational campaigns,
and in a relative sense with respect to MACHO lensing.
In fact, the EROS collaboration discussed the MACHO fraction
out of a single candidate event observed towards the SMC
and none towards the LMC, and the observed rate
of the recent OGLE-II and OGLE-III
campaigns is of the order of a few events only.
This is in fact a bonus as the small
``background'' signal of self-lensing events
with respect to MACHO lensing facilitates
the statistical interpretation of the results.
However, the small statistics of events at disposal
becomes a problem to the extent that it
makes difficult an analysis
of the event characteristics which
may shed more light on the nature of the events
beyond the ``simple'' statistics based essentially on 
the \emph{number} of events alone
(as outlined, eg, in the analyses of Mancini et~al, 2004 
and Calchi~Novati et~al, 2006 where
the event duration and spatial distributions
were used to this purpose). 
A larger self-lensing statistics would therefore definitively
help as it would make self-lensing signal
a robust test case so to finally better probe MACHO lensing.
The relevance of carrying out more detailed analyses on the
lens nature is made apparent also by the thorough
analysis of OGLE-2005-SMC-001 (Dong et~al, 2007),
out of which the authors conclude
that ``halo lenses are strongly favored but
SMC lenses are not definitively ruled out''.
In fact, Dong et~al (2007) conclusions are at odds
also with respect to the results of the MACHO
collaboration as they favor, to explain the lens, 
a Milky Way halo $\sim 10~\mathrm{M}_\odot$  \emph{black hole} solution,
namely in a mass range already excluded by Alcock et~al (2000a).
This composite picture strongly motivates the
still ongoing observational effort towards the LMC and the SMC carried out
by OGLE-IV (Udalski, 2011) and MOA-II (Sumi, 2011).

\section{Pixel Lensing towards M31}
Lying at about 770~kpc, the nearby giant spiral galaxy of M31 is more than one
order of magnitude more distant than the LMC
(with a difference in distance modulus between the LMC and M31 of about 6).
This makes the potential sources of microlensing
events along this line of sight \emph{unresolved}
for most ground-based telescope. This is the
underlying reason for the different regime
one enters in that case, the so-called
``pixel lensing'' (Gould, 1996). The source
being unresolved, the source flux
is usually more difficult, if not impossible, 
to be estimated from the light curve data alone.
This complicates the analysis as it adds an additional degeneracy in the
microlensing parameter space (M31 is however, fortunately,
near enough so that light curve data, if sufficiently
well sampled, especially along the wings of the microlensing
flux variations, may allow one to break this degeneracy). On the other hand,
with respect to the LMC, there is a much larger
number of potential sources available within
a comparatively much smaller cone of view. (In fact,
the sensitivity of past and current observational campaigns
allowed to detect only microlensing flux variations
out of very bright sources lying at the bright end
of the luminosity function, roughly stars brighter
than $M_I\sim 2$. Since this is a magnitude range where
the luminosity function is still undergoing an extremely
steep rise, an increase in sensitivity may therefore
bring to hugely extend the number of potential sources
and accordingly of the expected signal.) M31 is therefore
a target of choice for the search of MACHOs. A main
bonus of this line of sight is the possibility 
to map all of the M31 dark matter halo, which is not possible for the
Milky Way one. In fact, at parity of halo mass fraction
and MACHO mass, one expects about one third of MACHO events
from the Milky Way dark matter halo. This is because
the (much) larger number of available lenses in M31
is (almost) exactly compensated by the (much) larger
cross section of the Milky Way halo lenses.

The observational results towards M31 have been reviewed in Calchi~Novati (2010).
We recall the 4 years (1999-2002) observational campaign carried out
at the 2.5m INT telescope, whose data have been
analysed, in a completely independent way, by
the POINT-AGAPE and the MEGA collaborations.
Interestingly, the conclusions reached by these two 
groups on the MACHO issue differ. For  POINT-AGAPE, Calchi~Novati et~al (2005) reported
an evidence of a MACHO signal along this line of sight,
with in particular a \emph{lower} limit for $f$ of about 20\% in the mass
range $(0.1-1)~\mathrm{M}_\odot$. This result has been however challenged
by MEGA (de Jong et~al, 2006) who concluded that self lensing could
explain the full observed rate and placed an \emph{upper}
limit on $f$ of about 30\% in the same mass range.
The small statistics is an issue
for the POINT-AGAPE analysis which is based on 5 reported microlensing candidates
and for which, in addition, the robustness of the conclusions on the MACHO issue 
rests mainly upon a single candidate, PA-99-N2, for which,
according to the POINT-AGAPE analysis, the 
large distance from the M31 centre makes very unlikely
the attribution to a self-lensing population. 
The analysis of the MEGA collaboration, based on a more than twice as much larger set
of microlensing candidates, has been on the other
hand challenged on the basis that the characteristics
of the observed events are in fact at odds with
a full self-lensing explanation (Ingrosso et~al 2007).

Even beyond its importance within the POINT-AGAPE
analysis, PA-99-N2 is a very peculiar event.
The  good coverage  and tested achromaticity in three different bands
make of it a a very robust candidate in itself. 
A  caveat is the rather large full width at half maximum
duration, about 20~days (coupled
to a very large flux deviation at maximum), at the very limit  of the usually expected shorter 
values (a prediction confirmed by several other microlensing candidates).
The genuine microlensing nature of this flux variation
is on the other hand confirmed also by 
an anomaly along the light curve which turned out to
be compatible with a binary lens model (An et~al, 2004).
Indeed, it was subsequently observed that the secondary
lens could be due to a planetary system (Ingrosso et~al 2009, 2011)
although the large uncertainty on the model parameters
does not allow sharp conclusions on this issue.
(This would make of PA-99-N2 the second extra-galactic
exoplanet \emph{candidate} possibly ever observed after 
that of the microlensing event MACHO LMC-1,
the very first event reported by the MACHO collaboration
towards the LMC, Alcock et~al 1993, which remarkably
showed an anomaly that could even be explained
by an exoplanet secondary lens Dominik \& Hirshfeld 1994, 1996;
Alcock et al, 2000b).

The disagreement of the MEGA and the POINT-AGAPE analyses
may be traced back to two main lines of explanation,
both turning around the problem of 
the estimate of the expected signal. 
First, the evaluation of the expected self lensing signal,
due to the M31 luminous populations, still suffers
from possibly inaccurate modeling of these components.
Second, the correct evaluation of the \emph{efficiency} of a given pipeline,
the essential link to meaningfully compare the theoretical prediction
with the observed rate is, even more for M31
than for the LMC case, highly non trivial.

To complicate the analysis one has to face
the fact that the expected self lensing signal, at least
in the central M31 region,  is about
of the same order of  magnitude of an hypothetical
MACHO population in the  stellar mass range.
This complication comes together with the 
additional degeneracy of the lensing parameter space peculiar
to the pixel lensing regime which makes usually
difficult reliable estimates on the physical
time scale of the event and therefore
even more difficult any investigation on the lens \emph{nature}.
In fact the statistics of choice to disentangle
MACHO lensing from self lensing is the event spatial distribution, with 
self lensing expected to be, relatively, more clustered
around the M31 centre (in fact, whatever
the lens population, the signal gets larger towards the M31 centre
following the spatial density distribution of the sources,
at least as long it is not severely suppressed
by the huge background noise of the underlying
M31 surface brightness profile). 
As a second order effect, M31 MACHO lensing,
because of the inclination of M31 along the line
of sight, is expected to show an asymmetry
in the spatial distribution which should help
to characterize this signal (Crotts, 1992; Baillon et al 1993; Jetzer 1994).
The (recurrent) caveat
is the small statistics at disposal which has made difficult, at least up to now,
to efficiently carry out this kind of analyses.

Parallel to the statistical analysis of full data set, 
as those of POINT-AGAPE and MEGA,
the importance of carrying out thorough analyses of \emph{single} events
has become in the meantime increasingly clear.
Whenever they are possible, namely when the data are endowed with a sufficiently good
sampling along the flux variation, these analyses 
are extremely useful as they allow one to
extract additional and useful information
on the lens nature. This has probed to be the
case in particular for two events, both enjoying
coverage by two different data set. (Incidentally,
the routinely performed merging of
several data set for events observed
and followed up towards the Galactic bulge
should not lead to underestimate the 
importance of having successfully carried out a similar analysis
for M31 pixel lensing events.)
The PA-S3/GL1 event, first reported by the POINT-AGAPE collaboration 
(Paulin-Henriksson et~al, 2003)
and then by the WeCAPP collaboration (Riffeser et~al, 2003)
and subsequently analysed in more detail by the Riffeser et~al (2008).
Based on the extremely large
flux deviation at maximum amplification the authors
of this last work conclude, even for an event lying in the very inner
M31 region, strongly in favor of a MACHO hypothesis for the lens nature.
A similar conclusion, even if with a lower statistical significance, was reached
in the analysis of the OAB-N2 event carried out
by the PLAN collaboration (Calchi~Novati et~al 2010).
In this case the analysis was based on the
study of the finite size source effect
and the related study of the lens proper motion
made possible by the excellent sampling along
the full microlensing flux variation
together with a careful analysis on the 
unlensed source flux based on archive HST data.

More recently, Lee et~al (2011) presented the first results
of a new challenging observational M31 pixel
lensing project, PAndromeda, carried out at the 1.8m PS1
telescope with a huge field of view (7~deg$^2$). In particular
they report 6 new candidate events out
of their first season of data, for an analysis restricted to a subfield of $20'\times 20'$
around the M31 centre, showing in particular
excellent flux stability. New data and analyses from this
project are expected.

\subsection{The PLAN campaign at OAB\footnote{Results reported on behalf of the PLAN collaboration.}}
As a PLAN (Pixel Lensing Andromeda) collaboration
we have being carrying out a pixel lensing observational
campaign towards M31 using the 152cm Cassini telescope
located at ``Osservatorio Astronomico di Bologna''
in central Italy (at declination $44^{\circ}.4$, perfectly
suitable to observe M31).
Following a shorter pilot season in 2006, during 
the following 4 years PLAN has been allocated  
about 50 (almost) consecutive full nights/year.

\begin{table}[h]
\caption{\label{sampling} Statistics of the observational sampling for the PLAN campaign at OAB:
(1): year of observation; (2): number of good over allocated nights;
(3) average number of hours, per good night, of M31 observations.
}
\begin{center}
\begin{tabular}{cll}
\br
(1) & (2) & (3) \\
\mr
2006&  8/11 (73\%) & 4.2\\
2007& 31/50 (62\%) & 3.8\\
2008& 38/65 (58\%) & 4.6\\
2009& 25/36 (69\%) & 5.5\\
2010& 20/41 (49\%) & 4.6\\
\br
tot & 122/203 (60\%) & 4.5\\
\br
\end{tabular}
\end{center}
\end{table}

The observational strategy is set so to match
the characteristics of the expected signal
and maximise, at the same time, the expected rate.
Microlensing variations in M31 are typically expected
to last (in term of the full width at half maximum
duration, $t_{1/2}$, which characterize the observed signal)
a few days only. A regular sampling with high time-resolution
along the expected flux variations is therefore
essential first to distinguish microlensing variations
from other kind of intrinsic variable signals, second
to properly characterize the lensing parameter space
(as mentioned, a good enough sampling can even
allow one to break the degeneracy and get reliable
estimate of the physical duration, the Einstein time $t_\mathrm{E}$).
As a result, a long enough baseline (at least 2-3 weeks)
of consecutive nights is essential to carry out such
an observational programme. A longer baseline
is then useful both to increase the statistics of expected
events and to better constrain microlensing
against other variable signals.
Achromaticity is a further peculiar 
characteristic of microlensing 
useful to be exploited so that
observations should preferentially be carried out in two bands
(in particular, the expected sources being
typically luminous giants, we have used two broad-band filters similar to $R$, $I$ Cousin).
To reach a large enough signal-to-noise level is also essential
as we have to overcome the large background noise
set by the M31 surface brightness profile.
With a 1.5m class telescope we have
therefore to get to a long enough equivalent integration time per night,
at least of the order of 1-2 hours per filter.
With a CCD field of view of $13'\times 13'$
(pixel scale of $0.58''$), we have carried out
observations towards 2 fields
respectively North and South the M31 centre,
leaving out the inner 2 arcmin. This way
we cover the inner M31 part, in fact almost
all of the M31 bulge region, still we reach
far enough away from the M31 centre so to be able,
at least in principle, to disentangle 
MACHO lensing from self lensing on the basis
of the spatial distribution of the events.
Overall our observational strategy, requiring
about 8 hours per night (2 hours/field/filter), 
is set to match the number of potentially
available hours per night with our target
comfortably below an airmass of 1.5.
Full details on the  set up
and observational strategy are further 
discussed in Calchi~Novati et al (2007).

The statistics of the observations are reported
in Table 1. Overall, the average weather conditions
(humidity, cloud coverage)
have not shown to be optimal to our purposes,
introducing a lot of unwanted gaps in the sampling.
The ratio of at least partially
clear nights has indeed remained about the 60\% level,
which is barely sufficient, with however in addition 
an average number of hours we could actually observe M31
during those nights below 5 hours, so that the overall fraction
of hours useful to our purposes, with respect
to those allocated and potentially usable,
turned out to be in fact well below 40\%. 

The search for flux variations compatible with
a microlensing signal is carried out
with a fully \emph{automated} selection pipeline.
The result of this analysis are then compared
to the expected signal evaluated through
a full Monte Carlo simulation of the experiment,
where the observational set up is reproduced
together with the astrophysical and the microlensing
amplification model (in particular, we duly include
the finite size source effect).
The events \emph{selected} within the Monte Carlo
are then simulated on the \emph{raw} data upon
which we apply from scratch the  selection pipeline
(we stress that in this last step the simulation
is being carried out on the \emph{images}).
This way we get to an extremely reliable
estimate of the \emph{efficiency} of the selection pipeline
which in turn allows us a correct estimate
(for a given astrophysical model) of the number of expected events.

Full details of the selection process, the Monte
Carlo simulation and the related efficiency simulation
as well as the adopted astrophysical model
are given in Calchi~Novati et~al (2009) where
we  have also reported the results of
the 2007 season of data. In particular, two 
flux variations passed the full selection pipeline
and accordingly have been classified as 
microlensing \emph{candidates}: OAB-N1 and OAB-N2. 
Although primarily made to the purpose
of maximising the efficiency (and the observed rate),
the selection pipeline is also intended to leave
the final sample of candidates completely free from
contaminations of intrinsic variable sources
(the only viable option so to draw some meaningful
conclusion when dealing with
a signal of \emph{a few} expected events compared
to \emph{thousands} of detected flux variations).
Still, with few exceptions, the fate
of microlensing events is to keep their
attribution of \emph{candidates} if not for being
altogether rejected. In this specific case,
both detected flux variations had some features
that did not make them appealing.
As a main problem, OAB-N1 is affected by noisy data,
but this could in fact simply reflect the overall
quality of our data set, and indeed
we could not find any  plausible alternative
explanation to microlensing (and its  estimated 
duration is well in agreement with the expected signal).
OAB-N2 is a high signal-to-noise flux variation
nicely following the microlensing shape.
However, as presented in Calchi~Novati et al (2009),
it  completely lacks data points along the descent. 
The genuine microlensing nature of OAB-N2
was to be strongly supported in a second moment when additional data,
in particular useful to properly sample the descent, 
were kindly made available to us
by the WeCAPP collaboration (Calchi~Novati et~al 2010).

The estimate of the expected signal resulted
in about 1 self-lensing event,
and about the same for \emph{full} haloes
(both M31 and the Milky Way) of $0.5~\mathrm{M}_\odot$ MACHOs
(and about twice as much for $10^{-2}~\mathrm{M}_\odot$ MACHOs).
Comparing to the two candidate microlensing events
the observed rate looks therefore in agreement with
the expected self-lensing rate, the expected MACHO signal
on the other hand being too small to allow us to draw
meaningful conclusions.

The perspective on this initial tentative result was to be modified by the already
reported  thorough analysis of OAB-N2 (once the sampling completed by the WeCAPP data) 
which allowed us to conclude that OAB-N2 should be, though not with very large significance,
attributed to MACHO lensing rather than to self lensing (Calchi~Novati et~al 2010).
We are currently completing the analysis of the remaining three years of data
(Calchi~Novati et~al, 2012). Hopefully, the enlarged statistics is going to allow
us to draw sharper conclusions on the candidates lens nature and the MACHO issue.

\section{Conclusions}

Nowadays the effort of the observational microlensing community
is mainly driven by the search of exoplanets. The search of 
dark matter at the galactic scale in the form of compact halo
objects remains however an important and challenging
open issue. In fact, this second line of research is susceptible
to benefit from the exoplanet effort and this opportunity
should not be lost. The upgraded 
OGLE-IV and MOA-II systems, even with their main target 
being the Galactic bulge, will give the chance to proceed
to new and more thorough researches of MACHOs towards
the Magellanic Clouds. As for M31, a main challenge is
to enlarge the observed rate up to a level for which
first, the MACHO issue could be fully addressed 
on the basis of a large enough statistics.
Second, a large enough rate would allow parallel lines of research to be addressed, as those
carried out towards the Galactic bulge, first of all
the search of exoplanets
(lines of research for which microlensing would
constitute one of the few, if not the only, available tool).
A further challenge is to get, for M31 as already currently
done for the Galactic bulge targets, to a survey alert system 
coupled to a follow up round-the-clock coverage of the ongoing events.
To this purpose, however, one has to face
the difficulty linked to acknowledge
the microlensing nature of a pixel lensing
variation already during its raising part
also related to the usually rather short duration 
of the flux variations involved. 
This remains a compelling step 
to fully address the issue of the correct understanding
of the lens parameter space.

Microlensing observational campaigns have
robustly established that compact halo objects 
\emph{do not} constitute the main fraction
of galactic dark matter haloes. The picture
is still to be fully understood, however, in particular for
MACHO in the mass range $(0.1-1)~\mathrm{M}_\odot$:
a sizeable fraction of these objects, as that suggested by some results,
would indeed in any case represent a relevant challenge 
to our understanding of galactic astrophysics.
The larger statistics of events expected
from the ongoing campaigns towards both the
Magellanic Clouds and M31 promise to help
to finally solve the puzzle.

\section*{References}
\begin{thereferences}
\item 
Alcock C et~al 2000a, \emph{ApJ} {\bf 542}, 281
\item 
Alcock C et~al 2000b, \emph{ApJ} {\bf 541}, 270
\item 
Alcock C et~al 1993, \emph{Nature} {\bf 365}, 621
\item
An J et~al 2004, \emph{ApJ} {\bf 601}, 845
\item
Baillon~P, Bouquet~A, Giraud-H\'eraud~Y and Kaplan~J 1993,
\emph{A\&A} {\bf 277}, 1
\item
Bennett D 2005, \emph{ApJ} {\bf 633}, 906
\item
Calchi~Novati S 2010, \emph{GRG} {\bf 42}, 2101
\item
Calchi~Novati S et al 2012, \emph{in preparation}
\item
Calchi~Novati S and Mancini L 2011, \emph{MNRAS}  {\bf 416}, 1292
\item
Calchi~Novati S et al 2010, \emph{ApJ}  {\bf 717}, 987
\item
Calchi~Novati S et al 2009, \emph{ApJ}  {\bf 695}, 442
\item
Calchi~Novati~S, de~Luca~F, Jetzer~Ph, Mancini~L and 
Scarpetta~G 2008, \emph{A\&A} {\bf 480}, 723
\item
Calchi~Novati S et al 2007, \emph{A\&A} {\bf 469}, 115
\item
Calchi~Novati S, de~Luca~F, Jetzer~Ph,  and 
Scarpetta~G 2006, \emph{A\&A} {\bf 459}, 407
\item
Calchi~Novati et~al 2005, \emph{A\&A} {\bf 443}, 911
\item
Crotts A 1992, \emph{ApJ} {\bf 399}, L43
\item
de Jong et al 2006, \emph{A\&A} {\bf 446}, 855
\item
Dominik M 2010, \emph{GRG} {\bf 42}, 2075
\item
Dominik M et~al 2010, \emph{AN} {\bf 331}, 671
\item
Dominik M and Hirshfeld  A 1996, \emph{A\&A}  {\bf 313}, 841
\item
Dominik M and Hirshfeld  A 1994, \emph{A\&A}  {\bf 289}, L31
\item
Dong S 2007, \emph{ApJ} {\bf 664}, 862
\item
Gaudi S 2010,    Refereed chapter in EXOPLANETS, edited by S. Seager, 
to be published as part of the Space Science Series of the University of Arizona Press (Tucson, AZ),
(\emph{Preprint} arXiv:1002.0332)
\item
Gould A 2001, \emph{PASP} {\bf 113}, 903
\item
Gould A 2000, \emph{ApJ} {\bf 535}, 928
\item
Gould A 1996, \emph{ApJ} {\bf 470}, 201
\item
Gould A et al 2010, \emph{ApJ} {\bf 720}, 1073
\item
Ingrosso~G, Calchi~Novati~S, de~Paolis~F, Jetzer~Ph, Nucita~A and Zakharov~A
2011, \emph{GRG} {\bf 43}, 1047
\item
Ingrosso~G, Calchi~Novati~S, de~Paolis~F, Jetzer~Ph, Nucita~A and Zakharov~A
2009, \emph{MNRAS} {\bf 399}, 219
\item
Ingrosso~G, Calchi~Novati~S, de Paolis~F, Jetzer~Ph, Nucita~A, Scarpetta~G
and Strafella~F, 2007, \emph{A\&A} {\bf 462}, 895
\item
Jetzer Ph 1994, \emph{A\&A} {\bf 286}, 426
\item
Lee C.-H. et~al 2011, \emph{submitted to AJ} (\emph{Preprint} arXiv:1109.6320) 
\item
Mancini~L, Calchi~Novati S, Jetzer~Ph and Scarpetta~G 2004,
\emph{A\&A} {\bf 427}, 61
\item
Moniez M 2010, \emph{GRG} {\bf 42}, 2047
\item
Paczynski B 1986, \emph{ApJ} {\bf 304}, 1
\item
Paulin-Henriksson S et~al 2003, \emph{A\&A} {\bf 405}, 15
\item
Riffeser~A, Seitz~S and Bender~R 2008, \emph{ApJ} {\bf 684}, 1093
\item
Riffeser~A, Fliri~J, Bender~R, Seitz~S and G{\" o}ssl~C 2003, \emph{ApJ} {\bf 599}, L17
\item
Sumi T 2011, in XV International Conference on Gravitational Microlensing: Conference Book,
Bozza~V, Calchi~Novati~S, Mancini~L and Scarpetta~G eds (\emph{Preprint} arXiv:1102.0452)
\item
Sumi T {et~al} 2011, \emph{Nature} {\bf 473}, 349
\item
Sumi T {et~al} 2010, \emph{ApJ} {\bf 710}, 1641
\item
Tisserand P 2007, \emph{A\&A} {\bf 469}, 387
\item
Udalski A 2011, in XV International Conference on Gravitational Microlensing: Conference Book,
Bozza~V, Calchi~Novati~S, Mancini~L and Scarpetta~G eds (\emph{Preprint} arXiv:1102.0452)
\item
Wyrzykowski {\L} et~al 2011a, \emph{MNRAS} {\bf 416}, 2949
\item
Wyrzykowski {\L} et~al 2011b, \emph{MNRAS} {\bf 413}, 493
\item
Wyrzykowski {\L} et~al 2010, \emph{MNRAS} {\bf 407}, 189
\item
Wyrzykowski {\L} et~al 2009, \emph{MNRAS} {\bf 397}, 1228
\end{thereferences}

\end{document}